# Generalization of the Einstein-Plank-Richardson law for the photon energy in medium resolves Abraham-Minkowski dilemma in the electromagnetic field theory statement

Sergey G. Chefranov

A.M. Obukhov Institute of Atmospheric Physics, RAS, Moscow, Russia;

schefranov@mail.ru

**Abstract**

On the base of the Hamilton theory for the time-like photon in isotropic dielectric with refraction index n (S.Antoci, et.al, 2007), we suggest generalization of the Einstein-Plank-Richardson law for the value of the light energy quantum in medium: $E = h\nu n^2$, where h is the Plank's constant, and ν is the light frequency. By use of this new quantum law, we resolve the famous contradiction between de Broglie and Einstein's theories, related with the old Abraham-Minkowski dilemma in the definition of the photon momentum value p (in the medium for n>1). We show that the same value $p = p_a = \frac{E}{cn}$ (c is the speed of light in vacuum) follows now from the both theories of de Broglie and Einstein, which complies with the theory of Abraham, but not with the theory of Minkowski (where $p = p_m = \frac{En}{c}$). Based on the corpuscular approach with $p = p_a$ and $E = h\nu n^2$, we give new inference for the Snellius refraction law and resolve more old corresponding corpuscular-wave Newton-Huygens dilemma. We show that even for $n - 1 \ll 1$ the Abraham and Minkowski theories (with different $p = p_a$ and $p = p_m$) may lead to very different conclusions. Thus, only the theory with $p = p_a$, contrary to the theory with $p = p_m$, allows the conclusion about the Vavilov-Cherenkov radiation realization in the photon gas of the background cosmic radiation, for which $n^2 - 1 \cong 10^{-42}$ in the current epoch.

PACS: 03.50.De, 41.60.Bq, 98.70.Vc

# Generalization of the Einstein-Plank-Richardson law for the photon energy in medium resolves Abraham-Minkowski dilemma in the electromagnetic field theory statement

Sergey G. Chefranov

## § 1. Introduction

1. The problem of selection of adequate theory of electromagnetic field (EMF) in the medium continues to be actual already more than a century starting from the famous opposition of EMF theories of H. Minkowski [1] and M. Abraham [2]. Really EMF exists not in the pure vacuum, where it is described by the Maxwell theory, but always in the some type of medium with non unit refraction index n. Even photon gas of the background cosmic radiation (BCR), having in the modern epoch value of n with $n^2 - 1 \cong 10^{-42}$, may play the role of such a medium, creating quanta of EMF when it interacts with sufficiently fast charged particles of the cosmic rays (see farther §4).

It was necessary about 60 years before the resolving of Abraham-Minkowski theoretical dilemma [1, 2] was obtained on the base of series of direct experiments in 1968-1977 years [3-6] (see also more later works [7]). In these experiments, Abraham's theory has got sufficiently rigorous quantitative support[1], and Minkowski's theory vice versa was definitely rejected (at least, for the case of relatively low frequency EMF investigated in [3-6]).

In spite of this conclusion [3-6], equal consideration of the theories [1, 2] is preserved up to now in [8-10] and others in relation with the actual complex problem of defining the value of photon momentum in the medium. More over, it is paradoxical, but it is very Minkowski theory and based on it theories (as quantum theory of Vavilov-Cherenkov radiation (VCR), created in 1940 [11], i.e. far before [3-6]) continue having very vast usage. Necessity for the revising of VCR theory [11], based on [1], by the way was long ago stated in [12]. However only in [13, 14] such revision is made and new quantum VCR theory based on the definition of the photon momentum in the media corresponding to Abraham's theory [2] is proposed.

---

[1] In [3-6], predicted on the base of theory [2] value of new density of macroscopic electro-dynamic force (complementing to the Lorenz force) is measured; its existence in the most general case is denied in theory [1].

2. Noted paradoxical state with the Minkowski EMF theory choosing is not compatible with conclusions of experiments [3-6]. It is in many aspects related with domination of subjective opinion of some authoritative proponents of Minkowski theory [1], who are refusing, in spite of elementary logics, its incorrectness (see [11, 15]). In this relation, in [16] it is pointed out on the logical controversy of such a position, since it is obvious that from two contradictory theories at least one shall be incorrect. Meanwhile in [11], it is stated (see p. 320 in [11]): «All mentioned allows to count tensor of Abraham being «correct», but as it appears to us, declaration of tensor of Minkowski as "incorrect" is possible only when approaching to the problem somehow formally. Really, in the majority of cases results obtained on the base of Abraham and Minkowski tensors are absolutely equivalent. This allows possibility in the corresponding cases not only of using the Minkowski tensor but even consider its usage fully reasonable, if some simplifications are gained in the result ». The main reason for such a support of theory [1], as it is noted in [11, 15], is related with application of theory [1] in constructing quantum VCR theory [11], in which authors of [11, 15] have not any doubts. At the same time, doubts in the VCR theory [11], stated first in [12], are also extended in [13, 14] by using more precise comparisons with experimental data on the VCR threshold (see table in [13, 14]). It is shown that VCR theory of V.L.Ginzburg (1940) [11], which gives the similar result to the classical stationary VCR theory of Tamm-Frank (1937) [11], can adequately describe only already stabilized VCR field in the medium, but not the threshold of VCR itself. The later is defined by the non-stationary process of emitting of VCR by medium when it lefts locally its equilibrium state due to the interaction with sufficiently fast moving charged particle. Also in [17], it is stated that any application of theory [1] may be justified only for describing already stabilized stationary processes in the medium with EMF participation. Only in this stationary limit the distinction in the tensors of energy-momentum of EMF in theories [1] and [2] is actually small [17]. However, in §4, we give an example, where it is shown that even in the limit of $n^2 - 1 \ll 1$ (when quantitative distinction of character values in theories [1] and [2] seems to be actually, according to [17], negligible) there exist important qualitative and quantitative differences in the conclusions of theories of VCR based on [1] and [2]. This points on the un-acceptance of neglecting of differences of theories [1] and [2] tendency to which may be found not only in [11], [15] (as it can be seen from the given above citation from [11]), but also in many modern works (see [17] and given therein references).

3. From the other side, there is an objective fundamental physical problem in the dilemma of theories [1, 2] in defining the value of photon momentum in medium. We mean not only complexity of the direct measurement of the momentum value p for an individual photon but mainly existence of the well-known contradiction between theories of Einstein and de Broglie in the defining of value of p [9, 18]. Meanwhile it is usually taken that from Einstein's principle of equivalence of mass and energy, it follows that $p = p_a = \dfrac{E}{cn}$, corresponding to the theory of Abraham [2], and from quantum-wave theory of de Broglie, the value of $p = p_m = \dfrac{En}{c}$, is obtained that complies with the conclusions of Minkowski theory [1].

In the present work, we show possibility (see §2) of the complete resolving of the mentioned above contradiction between theories of Einstein and de Broglie on the base of introducing and usage of the new form $E = h\nu n^2$ for light energy quantum in the medium with n>1. As a basis for this, we use obtained in [19] conclusion on invariance of value of $\dfrac{E}{\nu n^2}$ for time-like photon in Hamilton theory of quantum of EMF in isotropic dielectric without dispersion. We show possibility of compliance of the pointed value of quantum energy of photon in the media with the data of classical experiments on external photo effect.

We give (in § 3) a new inference of Snellius law based on the corpuscular approach using pointed E and $p=p_a$.

In § 4, we give an example of substantial difference of conclusions of VCR theories based on theories of Minkowski [1] and Abraham [2], even in the limit of $n^2 - 1 \ll 1$. Meanwhile, we show that only on the base of the theory [13, 14] (with p=p$_a$), VCR is possible in the photon gas of background cosmic radiation (BCR) in the modern epoch when $n^2 - 1 \cong 10^{-42}$ for such a medium.

## § 2. Momentum and Energy of the Photon in Medium

1. According to Einstein's principle of energy-mass equivalence it is possible to assess (see [9]) value of photon impulse $p = mV$ via its mass $m = E/c^2$ and speed v=c/n (in the medium without dispersion when phase and group velocities of light waves coincide). Meanwhile obviously we get value p=p$_a$, following Abraham's theory [2].

From the other side, usually it is counted (see [9, 18]), that from the quantum-wave de Broglie theory there must follow another representation $p = p_m = En/c$, corresponding to Minkowski theory [1] (see [11]). Momentum value p and de Broglie wave length $\lambda$ for photon are related by $\lambda = h/p$. Hence under given wave length $\lambda = c/n\nu$ ($\nu$ is light frequency), in the most general case, photon impulse is:

$$p = \frac{h\nu n}{c}. \qquad (1)$$

This very representation directly follows from de Broglie theory, but not the equality $p = p_m$. From (1), it can follow also representation $p = p_m$, only if additional condition that photon energy in the media E and in the vacuum E$_o$, coincide (i.e., when $E = E_o = h\nu$).

Assumption on equality $E = E_o = h\nu$ complies also with theory [19] (giving the generalizing of Hamilton's photon theory [20] for the case of isotropic dielectric without dispersion), where invariance of $\frac{E}{\nu}$, is shown, but only in the case of space-like photon. For such a photon, it is found out that effective rest mass m$_p$ is not zero, but it is complex (i.e. m$_p{}^2$ < 0), which is defined from relativistic relationship $m_p^2 c^4 = E^2 - p^2 c^2$, where E and p are energy and momentum of photon in the medium. In the case of Minkowski, $p = p_m$ and actually we get $m_p^2 < 0$ for n>1 [13, 21].

At the same time, for time-like photon with $m_p^2 > 0$, according to [19], already another value $\frac{E}{vn^2}$ is invariant for n>1. In Abraham theory for mass $m_p$ always $m_p^2 > 0$ since $m_p$ has representation [13, 14]:

$$m_p = \frac{E\sqrt{n^2-1}}{c^2 n} \qquad (2)$$

for n>1. For n<1, according to [2], value $p_a = \frac{En}{c}$ [11] and in (2), it is necessary simply to replace n by $\frac{1}{n}$ [13, 14]. Hence, for any $n \neq 1$ in theory [2], value $m_p$ is finite and real valued.

In the result, for n>1 according to [19] and [2] for time-like photon of Abraham's theory, energy E may already not coincide with vacuum value, but have the form:

$$E = h\nu n^2. \qquad (3)$$

Now, from (3) and (1), according to de Broglie theory, we also get representation p=$p_a$, exactly coinciding with the estimate p obtained from theory of Einstein and Abraham (see also below (7) in §5.3, which is coinciding with (3) when dispersion is absent).

Thus, under condition (3) only Abraham theory [2] is found to be in the exact compliance with conclusions of the both theories of Einstein and de Broglie when defining momentum value in the medium. So, dilemma of theories Abraham-Minkowski again is found to be resolved for the benefit of Abraham theory [2], as it already takes place on the base of experiments [3-6]. Actually, for the case of space-like photon corresponding to Minkowski theory [1], there is already no opportunity to agree conclusions of theories of Einstein and de Broglie in the estimate of momentum of such a photon.

Let us note also that in (3), value of ν corresponds to the light frequency just in the medium. If not to assume the possibility of equality ν and value of frequency of light in the vacuum $v_o$, then the representation (3) may be replaced by a more general relationship $E = \frac{v}{v_o} E_o n^2$, where $E_o = h v_o$ is the photon energy in the vacuum. Only for $v=v_o$, it follows from (3) that $E = E_o n^2$. It shall be taken into account when comparing (3) with the experimental data.

2. Known results of observation of external photo effect in the classical experiments by Ya. Kunts [22], D.V. Kornelius [23], R.A. Milliken, W.H. Souder [24] and O.V. Richardson. K.T. Kompton [25] do not exclude allowing of the relationship (3) and indirectly confirm it to some degree. Actually, in [24], it is shown that initially zero photo-sensitivity of a fresh-prepared surface of the metal (sodium) to the light with wave length $\lambda = 5461 A^o$ begins to rise in time passing to increase reaching substantial maximal value and only then monotonically tends to zero when time of observation is increasing. From the other side, when using the light with less wave length $\lambda = 2535 A^o$ the same fresh prepared surface is initially found to be photo-sensitive and with time passing this photo-sensitivity only monotonically decreases to zero. In [24], this presence of photo-sensitivity increase in the first case is explained as a result of existing on the metal surface of a film appeared due to interaction of remaining (after creation of depression by the vacuum device) active gases with the metal. Meanwhile, in [24], it is considered that matter of the film is more electro-positive than the pure metal itself. And photo-curve for $\lambda = 5461 A^o$ is a result of emergence and following destruction of the matter of the film. At the same time, for sufficiently short waves pure metal surface is already so strong photo-electrically active that its own curve of photo-current decreasing fully masks the pointed effect from the film on the metal surface. In [24], it is also made a conclusion on the necessity of revision of earlier obtained conclusions on the necessity of obligatory presence of some amount of gas contacting the metal for the possibility in general of an external photo-effect observation. According to [24], to get the photo-effect when increasing the degree of vacuum depressing, it is necessary only to use the light with shorter wave length.

If to consider results of [24] based on (3), then it may be supposed an opportunity of the light quantum energy increase with wave length $\lambda = 5461 A^o$ $n^2$ times due to the presence of the film with the refraction factor $n \approx 1,4677$. Meanwhile, value of n is found from the condition of equality of the value of $n^2$ to the ratio of the wave lengths $\lambda = 5461 A^o$ and $\lambda = 2535 A^o$.

Let us note also an interpretation of the results of experiments [22, 23] on the base of (3), where a quadratic law relating light quantum energy and frequency where $E \approx v^2$ (according to [26], e.g., data of work [23] better fit to the cubic dependency E on v, i.e. $E \approx v^3$). In the specified experiments distinctions from the linear dependency is exposed especially visible in the very long wave region of photo-effect realization. According to (3), it may correspond to the film effect, when also the refraction factor itself (when dispersion is present) is found to be dependent on the light frequency v (see in this case (7) from §5.3). Let us note also that according to [25], obtained according to measurements of the photo effect threshold evaluation of the value of Plank constant (see formula (20) in [25]) is equal to h=8,07*10$^{-27}$ erg.sec. and is found to exceed significantly the known value h=6,624*10$^{-27}$ erg.sec. If this distinction in the value of h to interpret on the base of (3), then for the value of film refraction factor in the experiment [25], we may get an estimate $n = 1,1$.

### § 3. Snellius Law

In the previous paragraph, it was shown that on the base of the new representation (3) for the light quantum energy, we can resolve corpuscular-wave dilemma related with the contradiction of theories of Einstein and de Broglie when defining photon momentum value in the medium. Meanwhile, Abraham theory [2] gets support not only from the side of direct experiments [3-6], but is found to be that very theory of EMF in the medium that, contrary to [1], is compatible at the same time with the both pointed out fundamental physical theories.

Let us show that on the base of (3) when p=p$_a$ there is an opportunity to eliminate even more old corpuscular-wave dilemma of Newton-Huygens related with the well known non-uniqueness of inference of Snellius law from corpuscular and wave approaches [27].

Thus, based on the wave principle of Huygens it may be elementary inferred (see [27]) the following known form of that law of light refraction for the light beam, falling from vacuum (where n=1), under the angle θ to the normal of the flat surface of the transparent dielectric with the refraction index n>1 :

$$S = \sin\theta / \sin\theta_1 = n, \qquad (4)$$

where $\theta_1$ is an angle formed by the light beam in relation to the normal when light spreads inside the dielectric.

From the other side, usually, when using Newton's corpuscular approach, one gets instead of (4) the following representation (see [27]):

$$S = p/p_o, \qquad (5)$$

where p is the value of the impulse of light particle (photon) in the di-electric, and $p_o$ is its value in the vacuum. Assuming that (see [27]) p=mv and $p_0=mv_0$, from (5) it follows that $S=v/v_0 < 1$ contrary to (4), where S>1 for n>1.

If now in (5), instead of pointed out above representation for p and $p_o$ to use expressions $p_o = Eo/c$ and $p = p_a = E/nc$, then from (5), one gets the value $S = E/nE_o$. It may exactly coincide with (4) only under condition of holding equality (3) (i.e. for $E = n^2 Eo$).

Let us note, that from (5), it also follows (4) in the case of the use of Minkowski $p = p_m$, if to use corresponding to the space-like photon (see [19]) equality $E = Eo$.

In that aspect, special importance is given to the conducting of the direct experiments to check the relationship (3), defining relation between the photon energy and the light frequency in the media with $n>1$. If the experiment confirms that $E \neq Eo$, then the basis will be given for the contradiction of the using the Minkowski form $p = p_m$ to the the Snellius law inference.

## § 4. Vavilov-Cherenkov Radiation in the Photon Gas

1. Let us show that widely spread opinion on the possibility of neglecting of distinctions of theories [1] and [2] (see e.g. citation from [11]) may be incorrect even in the limit of very small deviations of the value of n from unit when n>1.

Consider photon gas of background cosmic radiation (BCR) for which in the modern epoch $n^2 - 1 \approx 10^{-42}$ [28].

In [28, 29], it is considered the possibility of realization in the modern epoch of VCR effect in the photon gas of BCR due to its interaction with sufficiently fast $\upsilon > \upsilon_{th}$ relativistic charged particles of the cosmic rays. Meanwhile, in [29], conclusion is made about impossibility of VCR in the modern epoch basing on the standard expression for the threshold velocity $\upsilon_{th} = c/n$, given in the quantum VCR theory [11], based on the Minkowski theory [1]. Actually, for such a form of $\upsilon_{th}$ and pointed out above value of n, in [28], it is obtained that VCR is possible only for $\gamma > \gamma_{th} = 10^{21}$, where $\gamma = 1/\sqrt{1 - \upsilon^2/c^2}$. This estimate $\gamma_{th}$ contradicts to the known GZK cutting of cosmic ray particles energy spectrum [30, 31], that does not allow realization of values with $\gamma >= 10^{11}$. In [29], estimate $n - 1 \approx 10^{-48}$ was used that leads to the value of $\gamma_{th} \approx 10^{24}$ also contradicting to GZK cutting.

Use now taking into account finiteness of (2) estimate $\upsilon_{th} = c/n*$, ($n* = n + \sqrt{n^2 - 1}$, for $n \geq 1$, and $n* = \frac{(1 + \sqrt{1 - n^2})}{n}$ for $n < 1$), given in the based on theory [2], new quantum VCR theory [13, 14]. Let us estimate possibility of VCR in the photon gas BCR in the modern epoch (see also [32]). For the used in [28] estimate of the value of $n^2 - 1 \approx 10^{-42}$ one can get that VCR is possible for $\gamma > \gamma_{th} = 2*10^{10}$, that already does not contradict to GZK cutting and means possibility of VCR observing in the modern epoch.

2. To define intensity of such VCR effect, let us use the known representation [29]:

$$\frac{dN}{d\omega \, dl} = \frac{\alpha}{c(E_{th}/mc^2)^2}, \qquad (6)$$

where dN is the number of Cherenkov photons emitted in the range of frequencies dω on the way of the length dl. In (6), $E_{th}$ is the threshold energy of a charged particle defined by the value $\upsilon_{th}$ or $\gamma_{th}$ in the limit $E_{th} \gg mc^2$, where m is the mass of the particle, and $\alpha = e^2/\hbar c$ is the constant of the thin structure. Meanwhile, in [29], it is obtained for $n = 1 + \Delta n$ ($\Delta n \ll 1$), that $E_{th} = mc^2 \sqrt{\frac{1}{2\Delta n}}$ and for $\Delta n \cong 10^{-48}$ on the way of length L of order of 1 Mpc ($3*10^{24}$ см), according to (6), the number of Cherenkov photons is found to be only $N \leq 3*10^{-5}$, that is not available for observation. From the other side, for the estimate $E_{th} = \frac{mc^2}{(8\Delta n)^{\frac{1}{4}}}$ and the same values of $\Delta n$ and L, one can get already available for observation value $N \approx 3*10^{19}$ according to (6).

## § 5. Conclusions

1. In the previous paragraph, example is given in which usage of theories of Minkowski [1] and Abraham [2] for estimation of the possibility for VCR in the photon gas of BCR leads not only to the qualitatively different conclusions, but also to the estimates of the number of Cherenkov photons differing in the order of magnitude by 24 in the limit of small $n^2 - 1 \ll 1$. It means that in the general case, it is not possible to use Minkowski theory [1] hoping on noted in [11] (see above § 1) "equivalence" of results obtained from [1] and on the base of "correct" Abraham theory.

Returning to the considered in § 1 problem of selection of adequate EMF theory in the media, we can make a conclusion on the necessity of revising on the base of theory [2] of all theories using at any degree conclusions following from Minkowski theory [1]. Meanwhile, obtained in [17] conditions of applicability of Minkowski theory [1] can't exclude conclusions similar to those made in § 4, since they are based on [17] only for assessing of distinctions of values of EMF energy-impulse tensors in theories [1] and [2]. Meanwhile, "applicability" conditions [1] correspond in [17] to the condition of these distinctions to be small. In the given in § 4 example (being a counterexample for theory [17]) distinctions in the values of $p_m$ and $p_a$ have astronomical order of smallness in the considered case when $n^2 - 1 \cong 10^{-42}$. Nevertheless, conclusions following from the VCR theory based on Minkovski theory [1], and from theory [13, 14], based on [2], are found to be diameter controversial regarding the possibility of VCR in the photon gas of BCR in the modern epoch.

2. Let us note that in many works (see [17, 33] and references therein) it is pointed out above tendency to neglect distinctions of theories [1] and [2] because full description of the system EMF-medium necessarily must include also medium energy-momentum tensor (EMT). Meanwhile, e.g., in [33], due to fitting artificial medium EMT choice, actually they achieve exact coincidence of corresponding integral medium EMT and EMF of theory [1] with EMT for EMF of theory [2]. This procedure has given in particular possibility to give in [17] an estimate of the value of added members of such a "medium" defining quantitative distinction of EMT for EMF in theories [1] and [2]. No physical sense such an adjustment has. Contrary, it may be noted that theory [1] initially does not assume (see VCR theory [11]) taking into account of energetic characteristics of medium versus to theory [2], in which EMT is related only to description of EMF. Actually, new VCR theory [13, 14], based on theory [2], generally could not be built without such accounting of medium energy change $\Delta E = m_p c^2$ (where $m_p$ is defined in (2)), necessary for possibility of VCR emitting by the medium itself. The latter as it is known (see [34, 35]), defines the microscopic VCR mechanism (not considered in VCR theory of Tamm-Frank, nor quantum theory of Ginzburg [11]), differing it from various kind effects of bremstrahlung radiation directly by the charged particle.

3. In the conclusion, we note that existing difficulties (see, e.g. [36, 37]) in conducting and interpretation (up to now only few [38-42]) experiments on the defining of photon vector momentum value in the medium now are suggested to be replaced by relatively more simple experimental checking of the new representation (3) for the scalar light quantum energy value in the medium.

Here it is worth to note representation for E, given in [43] (see formula (14) in [43]):

$$E = \frac{c^2}{\upsilon_{ph}\upsilon}\hbar\omega, \qquad (7)$$

that also as (3) generalizes (and it is noted in [43]), formula of Einstein-Plank[2] for "photon in the media" [43]. In (7), where $\hbar = \frac{h}{2\pi}$, $\upsilon_{ph} = \frac{c}{n}$ is light phase velocity, v is the group velocity or photon velocity as a particle. When dispersion is absent, i.e. when $v_{ph}=v$, expression (7) exactly coincides with (3).

Representation (7) was got in [43] without any relation with theory [19], but based on the use of conditions of compliance of theories of de Broglie and Einstein when estimating the value of "effective" rest mass of photon $m_{eff}$, obtained by photon due to interaction with the medium. Thus, in [43] value of $m_{eff}$ is obtained (see formula (13) in [47]):

$$m_{eff} = \frac{\hbar\omega\sqrt{1-\frac{\upsilon^2}{c^2}}}{\upsilon_{ph}\upsilon}, \qquad (8)$$

That follows from the comparison of definitions of photon momentum values in de Broglie theory $p = \hbar k$ (for $k = \frac{n\omega}{c}$, $\omega = v2\pi$ and $\upsilon_{ph} = \frac{\omega}{k}$) and in the relativistic theory where $p = m_{etf}\frac{\upsilon}{\sqrt{1-\frac{\upsilon^2}{c^2}}}$. Relationship (7), in its turn is obtained in [43] from comparison of (8) and relativistic representation of the stream energy $E = \frac{m_{eff}c^2}{\sqrt{1-\frac{\upsilon^2}{c^2}}}$.

When dispersion is absent, i.e. when $\upsilon = \upsilon_{ph} = \frac{c}{n}$, it follows from (8) that $m_{eff} = m_p$, where the value of $m_p$ defined in (2), corresponds to the photon mass in the 14medium according to very Abraham theory [2]. It however has not drawn attention in [43] where erroneous conclusion is made on the correspondence of representation (7) (or (14) in [43]) to the photon momentum in Minkowski theory but not to the Abraham theory (as it follows from [19] in the form (3) when $v_{ph}=v$). Pointed out contradiction is accepted in [43] because of a priori application of equality $E=E_o$ used in [43] for defining of photon momentum in the medium. So, in [43], it is defined $p_m = p_{free} n$ for Minkowski theory and $p_A = \frac{p_{free}}{n}$ for Abraham theory. Here $p_{free} = \frac{E_{free}}{c}$ and it is assumed that $E_{free} = E_o$. Meanwhile in [43] $E_{free}$ coincides with photon energy in vacuum $E_o$, but not with the energy $E$ of photon in the medium which according to (3) and (7) may differ from $E_o$ for time-like photon with $m_p$ from (2).

---

[2] This formula of linear relation between photon energy and light frequency is more fair to call as Einstein-Plank-Richardson formula (see[25] and references therein on the earlier works of O.V. Richardson on this subject).

It is interesting to develop generalization of theory [19] for the case of accounting dispersion effects and conducting comparisons (of obtained from (7) dependency $E = \hbar\omega(n^2 + n\frac{dn}{d\omega}\omega))$ with experimental data making more accurate conclusions of the discussed above in § 2 experiments [22-25].

Once again, we note an importance of establishing in the experiment of the real dependency of photon energy in the medium not only on the corresponding light frequency but on the value of refraction factor n as well.